\documentstyle[12pt]{article}
\newcommand{\be}{\begin{equation}}
\newcommand{\ee}{\end{equation}}
\newcommand{\ba}{\begin{eqnarray}}
\newcommand{\ea}{\end{eqnarray}}

\begin{document}
\begin{center}
{\bf NEW METHODS FOR TWO-DIMENSIONAL SCHR\"ODINGER EQUATION:\\
$SUSY-$SEPARATION OF VARIABLES AND SHAPE INVARIANCE}\\
\vspace{0.5cm}
{\large \bf F. Cannata$^{1,}$\footnote{E-mail: cannata@bo.infn.it},
M.V. Ioffe$^{2,}$\footnote{E-mail: m.ioffe@pobox.spbu.ru} ,
D.N. Nishnianidze$^{2,3,}$\footnote{E-mail: nishnianidze@c4.com}}\\
\vspace{0.2cm}
$^1$ Dipartimento di Fisica and INFN, Via Irnerio 46, 40126 Bologna, Italy.\\
$^2$ Department of Theoretical Physics, Sankt-Petersburg State University,\\
198504 Sankt-Petersburg, Russia\\
$^3$ Kutaisi Polytechnical University, 384000 Kutaisi,
Republic of Georgia
\end{center}
\vspace{0.2cm}
\hspace*{0.5in}
\hspace*{0.5in}
\begin{minipage}{5.0in}
{\small
Two new methods for investigation of two-dimensional quantum systems, whose
Hamiltonians are
not amenable to separation of variables, are proposed.
1)The first one - $SUSY-$
separation of variables - is based on the intertwining relations of Higher order
SUSY Quantum Mechanics (HSUSY QM)
with supercharges allowing for separation of
variables. 2)The second one is a generalization of shape invariance.
While in one dimension shape invariance
allows to solve algebraically a class of (exactly solvable) quantum
problems, its generalization to higher dimensions has not been
yet explored. Here we provide a formal framework in HSUSY QM for two-dimensional
quantum mechanical systems for which shape invariance holds. Given the
knowledge of one eigenvalue and eigenfunction,
shape invariance allows to construct a chain of new eigenfunctions and
eigenvalues. These methods are applied to a two-dimensional quantum system,
and partial explicit solvability is achieved in the sense that only
part of the spectrum is found analytically and a limited set of
eigenfunctions is constructed explicitly.\\
\vspace*{0.1cm}
PACS numbers: 03.65.-w, 03.65.Fd, 11.30.Pb
}
\end{minipage}
\vspace*{0.2cm}
\section*{\bf 1. \quad Introduction}
\vspace*{0.5cm}
\hspace*{3ex}In one-dimensional Quantum Mechanics the importance of
exactly solvable and
quasi exactly solvable (QES) models has been stressed repeatedly. The approach
of Supersymmetric Quantum Mechanics (SUSY QM) and in particular shape
invariance \cite{review} has
been fully exploited for construction and investigation of
such models by generating a partnership between pairs of dynamical systems
which allows to establish the solvability of one in terms of the other
by means of intertwining relations with supercharges of first order in derivatives.

With this knowledge one can construct a variety of multidimensional
solvable quantum systems by suitable combination of solvable
one-dimensional dynamical systems (separation of variables and
its generalizations \cite{miller}). From now on we will focus the main
attention to
two-dimensional quantum systems without assuming that such separation of
variables is possible.
In two-dimensional Quantum Mechanics solvable (or partially solvable)
dynamical systems, for which the entire spectrum (or part of it) and the
associated wave functions are known, play a role similar to solvable (or QES)
models in one dimension.

Within the search for a larger class of problems, which can be solved
by supersymmetrical methods, extensions of SUSY QM have been elaborated
with different realizations of the intertwining operators (supercharges).
In particular, one-dimensional supercharges were constructed in terms of
higher-derivative operators \cite{HSUSY}, \cite{ourshape} (HSUSY QM),
the associated superalgebra
(HSUSY) containing a higher order polynomial of the Hamiltonian. This
generalization has been revisited
recently in \cite{aoyama}, \cite{chili}, \cite{sasaki} and referred
to as $N$-fold SUSY. Its most simplified version \cite{aoyama}
($A$-type $N$-fold SUSY)
corresponds to a solvable ansatz for the so called
\cite{HSUSY}
reducible supercharges and for more general factorized
supercharges.
However also a class of higher order supertransformations
was found \cite{HSUSY},
which can not be represented as a succession of two standard first order
supertransformations (the so called, irreducible supercharges).
Accordingly, the
second order intertwining operators were exhaustively classified as
reducible and irreducible. We warn that our definition of reducibility
\cite{HSUSY} does not coincide with factorization of supercharge.

An important step for investigation of one-dimensional HSUSY models
is the construction
of zero modes of  (higher order) supercharges which
can be instrumental for quasi exactly solvability \cite{aoyama},
\cite{sasaki} and for the study \cite{ourshape} of spectral properties
of higher order
shape invariant systems. Indeed, it is well known that
in one-dimensional standard SUSY QM a very elegant method of
solution of the spectral
problem exists for potentials which preserve their shape in the SUSY
partnership \cite{review},\cite{shape}. It was shown that for a wide
variety of such potentials the
entire spectrum and the eigenfunctions can be obtained algebraically
providing a fresh reformulation of the old Factorization
Method \cite{infeld} for the Schr\"odinger
equation. Extensions of shape invariance to third order
not fully reducible HSUSY QM models were proposed in
\cite{ourshape}.

In the $N-$ dimensional SUSY QM \cite{ABEI} starting from a scalar
Schr\"odinger
Hamiltonian a chain of matrix Hamiltonians of different
dimensionality was constructed
ending with another scalar Hamiltonian.
Each pair of neighbouring Hamiltonians is intertwined
by a first order supercharges, each Hamiltonian has a partial
isospectrality with
both neighbours. In the two-dimensional case this chain simplifies
to two scalar Hamiltonians
and one matrix $2 \times 2$ Hamiltonian. The spectra of
the two scalar Hamiltonians
build up the spectrum of the matrix Hamiltonian, but in general
the spectra of the scalar
Hamiltonians are not related.

However, cases were found \cite{david}, \cite{classical},
where the two scalar two-dimensional
Hamiltonians
are intertwined by second order supercharges and are therefore isospectral,
up to
zero modes of the supercharges. This suggests that HSUSY QM may even be
more important
for the two-dimensional case than in one-dimensional case
to study spectra and
eigenfunctions of two-dimensional models not amenable to
separation of variables.
Due to the complexity of the system of nonlinear
partial differential equations,
arising from the second order intertwinig relations,
one has to look only for particular
solutions. Indeed, two classes of such particular systems were found in
\cite{david}, \cite{classical}.

In Section 2 we outline the second order SUSY QM framework in one and two
dimensions, which allows to formulate in Section 3 the method of $SUSY-$
separation of variables for two-dimensional Hamiltonians, which are not
amenable to separation of variables. This method can be used when the
{\it supercharges} allow for separation of variables. In Section 4 a
two-dimensional singular Morse type potential (Subsection 4.1)
is shown to satisfy
$SUSY-$separation of variables. For this model the zero modes of the supercharge
(Subsection 4.2) and eigenfunctions of the Hamiltonian (Subsection 4.5)
in the linear space of zero modes of the supercharge are constructed. In addition,
a new eigenfunction outside this space is built explicitly (Subsection 4.6).
 The role of shape invariance (Subsection 5.1) has not been explored yet in the two-dimensional
SUSY QM (Subsection 5.2).
It is one aim of this paper (Subsection 5.3) to find
Hamiltonians (without separation of variables),
which realize the two-dimensional generalization
of shape invariance. The model of Subsection 4.1 is found to satisfy
shape invariance.
It is shown that, due to the nontrivial space of zero modes of supercharges,
shape invariance does not allow a fully algebraic solution for the entire
spectrum, but only (partial explicit solvability) for part of the
spectrum
and for the corresponding wave functions. Section 6
contains a brief discussion of the  two-dimensional Morse type
model in another
region of parameter-values and of the integrability of this model,
by elucidating the action of quantum integrals of motion, commuting with the
Hamiltonian, on the constructed eigenfunctions.

\section*{\bf 2.\quad Second order SUSY in one and two dimensions}
\vspace*{0.5cm}
\hspace*{3ex}Let us provide notations and the main formulas for standard
 SUSY QM \cite{review}:
\be
\tilde H = Q^+Q^- = -\partial^2 + \tilde V(x);\quad
\tilde H \tilde\Psi_n(x) = E_n \tilde\Psi_n(x);\label{h1}
\ee
\be
H = Q^-Q^+ = -\partial^2 + V(x);\quad
H \Psi_n(x) = E_n \Psi_n(x);\label{h2}
\ee
\be
\tilde HQ^+ = Q^+H;\quad
Q^-\tilde H = HQ^-;\label{intertw}
\ee
\be
Q^+ = -\partial + W(x);\quad
Q^- = (Q^+)^{\dagger} = \partial + W(x);\label{superch1}
\ee
\be
\Psi_n(x) = Q^- \tilde\Psi_n(x);\quad
\tilde\Psi_n(x) = Q^+ \Psi_n(x),\label{psi}
\ee
where $\partial \equiv d/dx.$

The spectral equivalence (up to zero modes of $Q^{\pm}$) of
$\tilde H, H$ can be expressed via the superalgebra:
\ba
 \hat H =
\left( \begin{array}{cc}\tilde H&0\\
0& H
\end{array} \right); \qquad
 \hat Q^+ =(\hat Q^-)^{\dagger} = \left( \begin{array}{cc}
                        0&0\\
                        Q^-&0
                        \end{array} \right);\nonumber\\
  \{\hat Q^+,\hat Q^-\} = \hat H;\quad (\hat Q^+)^2 =(\hat Q^-)^2 = 0;
  \quad [\hat H ,\hat Q^{\pm}]
  =0.
                                         \label{strepr}
\ea

The second order generalization of the superalgebra incorporates
\cite{HSUSY} the most general intertwining operators of second
order in derivatives:
\be
Q^+ = \partial^2 - 2f(x) \partial + b(x);\quad Q^- =
(Q^+)^{\dagger}.\label{superch2}
\ee
Intertwining relations (\ref{intertw}) with supercharges (\ref{superch2}) lead
\cite{HSUSY} to the expressions of the potentials $V,\tilde V$ and the
supercharges $Q^{\pm}$ in
terms of the only real function $f(x):$
\ba
 &&\tilde V(x) = -2f'(x) + f(x)^2 + \frac{f''(x)}{2f(x)} -
 \biggl(\frac{f'(x)}{2f(x)}\biggr)^2 - \frac{d}{4f(x)^2} - a;\nonumber\\
 &&V(x)= 2f'(x) + f(x)^2 + \frac{f''(x)}{2f(x)} -
 \biggl(\frac{f'(x)}{2f(x)}\biggr)^2 - \frac{d}{4f(x)^2} - a;\nonumber\\
&&b(x) = - f'(x) + f(x)^2 - \frac{f''(x)}{2f(x)} +
 \biggl(\frac{f'(x)}{2f(x)}\biggr)^2 + \frac{d}{4f(x)^2},\nonumber
\ea
where $a$ and $d$ are arbitrary real constants. The case $d\leq 0$
was called \cite{HSUSY} reducible, since it can be interpreted in terms
of two successive first order supertransformations with real
superpotentials and an hermitian intermediate Hamiltonian (up to
a constant
real energy shift). The alternative case $d>0$ was called irreducible.

A two-dimensional generalization of standard SUSY QM was
proposed in \cite{ABEI}, where the Superhamiltonian includes
a chain of one matrix and two scalar Hamiltonians. Each scalar
Hamiltonian is separately intertwined with the matrix Hamiltonian,
but the spectra of the two scalar Hamiltonians are not related.

If one wants to relate directly \cite{david}, \cite{classical}
the spectra of two
scalar two-dimensional
Schr\"odinger operators via intertwining relations  analogous to (\ref{intertw}):
\ba
&&\tilde H(\vec x)Q^+ = Q^+H(\vec x), \quad
Q^-\tilde H(\vec x) = H(\vec x)Q^-,\label{intertw2}\\
&&H = -\triangle + V(\vec x);\quad \tilde H = -\triangle + \tilde V(\vec x)
\quad  \triangle \equiv \partial_1^2 + \partial_2^2,\quad
\partial_i \equiv \partial/\partial x_i,\nonumber
\ea
it is expedient to make use of a two-dimensional generalization
of (\ref{superch2}):
\be
Q^+ = (Q^-)^{\dagger} = g_{ik}{(\vec x)}\partial_i \partial_k +
C_i(\vec x)\partial_i + B(\vec x), \label{superch22}
\ee
where all coefficient functions are real.

It means that, up to zero modes of $ Q^{\pm}, $ the spectra of
$H, \tilde H$ coincide and their eigenfunctions are mutually connected:
$$
\Psi \sim Q^-\tilde\Psi,\quad  \tilde\Psi \sim Q^+\Psi.
$$

A classification of the dynamical systems
 requires first to determine the "metrics" in supercharges
(\ref{superch2}).
The Eq.(\ref{intertw2}) leads to a set of four equations for the metrics
$g_{ik}(\vec x),$
\be
\partial_ig_{kl} + \partial_kg_{il} + \partial_lg_{ik}=0; \quad
\{ i k l \} = \{ 1 1 1; 1 1 2; 1 2 2; 2 2 2 \}, \label{metrics}
\ee
which can be solved independently from the other ones:
$$
g_{11}
= a x_2^2 + a_1 x_2 + b_1;
\,
g_{22} = a x_1^2 +
a_2 x_1 + b_2;\, g_{12}
=-\frac{1}{2}(2 a x_1 x_2 +  a_1 x_1 +
 a_2 x_2) + b_3,
$$
with $a,\,a_i,\,b_i$ - constants.

For the particular case of the unit metrics
$ g_{ik}=\delta_{ik} $ the corresponding quantum systems
allow \cite{david} for the so called\cite{miller} R-separation of variables in
parabolic, elliptic or polar coordinate systems.

After having solved (\ref{metrics}), the intertwining
relations (\ref{intertw2})
are equivalent to the following system
of differential equations for potentials $V(\vec x),\, \tilde V(\vec x)$ and
coefficient functions $C_k(\vec x), B(\vec x)$:
\ba
&&\partial_i C_k + \partial_k C_i +
\triangle g_{ik} - (\tilde V - V)g_{ik} = 0;
\nonumber\\ &&\triangle C_i + 2\partial_i B +
2 g_{ik}\partial_k V - (\tilde V -
V)C_i=0;\label{system}\\ &&\triangle B +
g_{ik}\partial_k\partial_i V + C_i\partial_i V
- (\tilde V - V) B = 0.\nonumber
\ea

Because of the complexity of the equations above, only particular solutions
have been found in \cite{david}, \cite{classical} by making suitable anzatzes. In particular,
for the supercharges with {\it Lorentz metrics}
$(g_{ik} = diag(1,-1)):$
\be
Q^+ = (\partial_1^2 - \partial_2^2)
+  C_k \partial_k + B = 4\partial_+\partial_- +C_+\partial_- +
C_-\partial_+ + B, \label{ourq}
\ee
a solution of (\ref{system}) can be reduced \cite{david}, \cite{classical}
to a solution
of the system:
\ba
&&\partial_-(C_- F) =
-\partial_+(C_+ F);\label{first}\\
&&\partial_+^2 F = \partial_-^2 F,\label{second}
\ea
where $x_{\pm} \equiv x_1\pm x_2\quad
\partial_{\pm}=\partial / \partial x_{\pm} $ and $C_{\pm}$ depend only on
$ x_{\pm},$ respectively:
$$C_+ \equiv C_1 - C_2 \equiv C_+(x_+);\quad
C_- \equiv C_1 + C_2 \equiv C_-(x_-).
$$
The function
$ F, $ solution of (\ref{second}), is represented as a sum
$$ F=F_{1}(x_{+}+x_{-}) + F_{2}(x_{+}-x_{-}).$$
The potentials $ \tilde V(\vec x), V(\vec x) $ and the function $ B(\vec x) $
are expressed
in terms of $F_1(2x_1),\,F_2(2x_2)$ and $C_{\pm}(x_{\pm}),$
solutions of system (\ref{first}), (\ref{second}):
\ba
\tilde V&=&\frac{1}{2}(C_+' + C_-') + \frac{1}{8}(C_+^2 + C_-^2) +
\frac{1}{4}\bigl( F_2(x_+ -x_-) - F_1(x_+ + x_-)\biggr) ,\nonumber\\
V&=&-\frac{1}{2}(C_+' + C_-') + \frac{1}{8}(C_+^2 + C_-^2) +
\frac{1}{4}\bigl( F_2(x_+ -x_-) - F_1(x_+ + x_-)\biggr) ,\label{potential}\\
B&=&\frac{1}{4}\bigl( C_+ C_- + F_1(x_+ + x_-) + F_2(x_+ - x_-)\biggr) .
\label{functionB}
\ea

\section*{\bf 3.\quad $SUSY-$separation of variables: a construction
of new two-dimensional partially solvable models}
\vspace*{0.5cm}
\hspace*{3ex} We want to study two-dimensional Hamiltonians without separation
of variables within the SUSY approach. Our goal is to find a class of
Hamiltonians for which  part of the spectrum and of the eigenfunctions can be
found (partially solvable
systems). Our approach is based on the solution of the intertwining relations
(\ref{intertw2}) for partner potentials and supercharges followed by
the investigation of zero modes of the supercharges (\ref{superch2}).
Because, in principle,
there is no connection between separation of variables in the supercharges
and the Hamiltonians, one can look for the opportunity of partially solving
for the spectrum of a Hamiltonian (which does not allow the separation of
variables) via normalizable zero modes of supercharges which do allow such
separation.

More specifically we will start from the investigation
of zero modes of supercharges which do allow for separation
of variables
and solve the intertwining relations (\ref{intertw2}) obtaining Hamiltonians
which are not amenable to separation of variables. The algebraic method
(for partial solution) will be presented in this Section in its general form.
In the next Section the method will be applied to a specific type of
two-dimensional models.

Let us suppose that there are $(N+1)$ normalizable zero modes
$\Omega_n(\vec x),\,\,n=0,1,...,N $ of the supercharge $Q^+$:
\be
Q^+ \vec\Omega (\vec x) =0,\label{zero}
\ee
where $\vec\Omega (\vec x)$ is a column vector with components
$\Omega_n(\vec x).$ Acting by intertwining (\ref{intertw2}) onto $\vec\Omega
(\vec x)$ it is easy to realize that the space of zero modes is
closed under the action of $H:$
\be
H\vec\Omega (\vec x) = \hat C \vec\Omega (\vec x),
\label{matrixC}
\ee
where $\hat C \equiv ||c_{ik}||$ is a $c-$number $\vec x-$independent
real matrix. If the matrix
$\hat C$ can be diagonalized\footnote{If $\hat C$
can not be diagonalized, the algebraic method can be
applied if one can solve the less restrictive
matrix equation $\hat B\hat C = \hat\Lambda\hat B$.}
 by a real similarity transformation:
\be
 \hat B \hat C (\hat B)^{-1} = \hat\Lambda =
diag(\lambda_0,\lambda_1,...,\lambda_N),
\label{diag}
\ee
the problem
reduces to a standard algebraic task within the zero modes space:
\be
H (\hat B\vec\Omega (\vec x)) = \hat\Lambda(\hat B\vec\Omega (\vec x)).
\label{diagonal}
\ee

For attacking the problem it is expedient to eliminate the first order
derivative terms in the supercharge (\ref{ourq}) by a suitable similarity
transformation:
\ba
 q^+ &=& \exp(-\chi (\vec x)) \,Q^+\, \exp(+\chi (\vec x)) = \partial_1^2 -
\partial_2^2 + \frac{1}{4}(F_1(2x_1) + F_2(2x_2)),
\label{qtilde}\\
\chi (\vec x) &=& -\frac{1}{4}\bigl( \int C_+(x_+)dx_+ +
\int C_-(x_-)dx_- \biggr).
\label{chi}
\ea

We notice that $ q^+$ exhibits separation of variables: this is what we
mean by {\bf $SUSY-$separation of variables}, even if the components of
the Superhamiltonian do not admit such separation.
Zero modes of $ q^+ $ can be found as a linear superposition of products
of one dimensional wave functions $\eta_n (x_1),$ and $\rho_n(x_2),$
satisfying Schr\"odinger equations:
\ba
(-\partial_1^2 -\frac{1}{4}F_1(2x_1))\eta_n(x_1)=\epsilon_n\eta_n(x_1),
\nonumber\\
(-\partial_2^2 +\frac{1}{4}F_2(2x_2))\rho_n(x_2)=\epsilon_n\rho_n(x_2),
\label{etarho}
\ea
with $\epsilon_n$ - the separation constants.

In analogy to (\ref{qtilde}), one can define operators
\be
h \equiv \exp(-\chi (\vec x))\, H\,\exp(+\chi (\vec x))=
-\partial_1^2-\partial_2^2 +C_1(\vec x)\partial_1 -
C_2(\vec x)\partial_2 -\frac{1}{4} F_1(2x_1) +\frac{1}{4}F_2(2x_2),
\label{tildeh}
\ee
$\tilde h$ and
eigenfunctions of $q^+$:
\be
\omega_n(\vec x)= \exp(-\chi (\vec x))\cdot\Omega_n(\vec x),
\label{omegatilde}
\ee
keeping however in mind that the normalizability and orthogonality is not
preserved automatically due to non-unitarity of the similarity
transformation. The operators $h$ and $\tilde h$ should not be interpreted
literally as
Hamiltonians since they are non-hermitian (but have real spectrum)
\footnote{Incidentally, we remark that by an additional unitary transformation
one can generate analytically a class of two-dimensional non-trivial
Hamiltonians with complex potentials but
with real spectrum. Examples of two-dimensional Hamiltonians have
been discussed recently in \cite{bender}.}, and
are not factorized in $q^{\pm}$, like in
(\ref{h1}), (\ref{h2}).

Then using (\ref{etarho}) one can write:
\be
h\omega_n(\vec x)=[2\epsilon_n +C_1(\vec x)\partial_1
-C_2(\vec x)\partial_2]\omega_n(\vec x).
\label{eigen}
\ee
While from the equation above it is not manifest, we know however
from (\ref{matrixC}) that the space spanned by functions $\omega_n(\vec x)$
is closed under the action of $h.$ In the concrete model of next Section
this will be demonstrated in Subsection 4.4.

It is clear that, in contrast to (\ref{qtilde}) where variables are separated,
there is no separation
of variables for $h,$ which makes the two-dimensional dynamics not-trivially
reducible to one-dimensional dynamics. For this reason we refer
to this method
for partial solvability as to {\bf $SUSY-$separation of variables}.

\section*{\bf 4.\quad A new two-dimensional partially solvable
model.}
\vspace*{0.2cm}
\subsection*{\bf \quad 4.1.\quad Formulation of the model.}
\hspace*{3ex}
In this Section we apply the method of Section 3 to
a particular example, which can be interpreted as a suitable
two-dimensional generalization of Morse potential.
Among the solutions of the system of equations (\ref{first}) and (\ref{second})
found in \cite{classical} we focus attention on the particular case
already presented in \cite{classical}:
\ba
&&F_1(x) = k_1\biggl( \alpha_+\alpha_-\exp(\lambda x) +
\beta_+\beta_-\exp(-\lambda x)\biggr)
+ k_2\biggl( \alpha_+^2\alpha_-^2\exp(2\lambda x) +
\beta_+^2\beta_-^2\exp(-2\lambda x)\biggr) ;\nonumber\\
&&-F_2(x) = k_1\biggl( \alpha_+\beta_-\exp(\lambda x) +
\beta_+\alpha_-\exp(-\lambda x)\biggr) +
 k_2\biggl( \alpha_+^2\beta_-^2\exp(2\lambda x) +
\beta_+^2\alpha_-^2\exp(-2\lambda x)\biggr) ; \nonumber\\
&&C_{\pm} = \pm \frac{\alpha_{\pm}\exp(\lambda x_{\pm}) +
\beta_{\pm}\exp(-\lambda x_{\pm})}
{\lambda\biggl( \alpha_{\pm}\exp(\lambda x_{\pm}) -
\beta_{\pm}\exp(-\lambda x_{\pm})\biggr) }; \label{16}\\
&&V=
\frac{2\alpha_+\beta_+ (1 + 8\lambda^{2}) +
\alpha_{+}^{2}\exp(2\lambda x_{+}) + \beta_{+}^{2}\exp(-2\lambda x_{+})}
{8\lambda^{2}\biggl( \alpha_{+}\exp(\lambda x_{+}) -
\beta_{+}\exp(-\lambda x_+)\biggr)^{2}} +\nonumber\\&&
+\frac{2\alpha_-\beta_- (1 - 8\lambda^{2}) +
\alpha_{-}^{2}\exp(2\lambda x_{-}) + \beta_{-}^{2}\exp(-2\lambda x_{-})}
{8\lambda^{2}\biggl( \alpha_{-}\exp(\lambda x_{-}) -
\beta_{-}\exp(-\lambda x_-)\biggr)^{2}} -\nonumber\\&&
-\frac{1}{4}\biggl[k_{1}\biggl(\alpha_{+}\beta_{-}\exp(2\lambda x_{2}) +
\alpha_{-}\beta_{+}\exp(-2\lambda x_{2})\biggr) +
k_{2}\biggl(\alpha_{+}^{2}\beta_{-}^{2}\exp(4\lambda x_{2}) + \nonumber\\&&
+\alpha_{-}^{2}\beta_{+}^{2}\exp(-4\lambda x_{2})\biggr) +
k_{1}\biggl(\alpha_{+}\alpha_{-}\exp(2\lambda x_{1}) +
\beta_{+}\beta_{-}\exp(-2\lambda x_{1})\biggr) + \nonumber\\&&
+k_{2}\biggl(\alpha_{+}^{2}\alpha_{-}^{2}\exp(4\lambda x_{1}) +
\beta_{+}^{2}\beta_{-}^{2}\exp(-4\lambda x_{1})\biggr)\biggr].\label{24}
\ea
We will consider a specific case of
the
expressions for $C_{\pm}$ and $V$ by choosing
the parameters in (\ref{16}) and
(\ref{24}) so that $\beta_+=0, \alpha_-=\beta_-,
\lambda \equiv -\alpha / 2, \alpha > 0.$
It is evident that Eq.(\ref{first}) admits a more general solution
than (\ref{16}) because
one can introduce an additional multiplicative parameter
in $C_{\pm}$ in (\ref{16}).
The
nonlinear dependence of (\ref{potential}) on
$C_{\pm}(x_{\pm})$ leads to a nontrivial generalization of the
corresponding potential.

We then obtain:
\ba
C_+&=&4a\alpha;\quad C_-=4a\alpha\cdot\coth \frac{\alpha x_-}{2};
\label{cpm}\\
f_1(x_1)&\equiv & \frac{1}{4} F_1(2x_1)=-A\biggl(\exp(-2\alpha x_1) -
2 \exp(-\alpha x_1)\biggr);
\label{f1}\\
f_2(x_2)&\equiv & \frac{1}{4} F_2(2x_2)=
+A\biggl(\exp(-2\alpha x_2) - 2 \exp(-\alpha x_2)\biggr);
\label{f2}\\
\tilde V(\vec x)&=& \alpha^2a(2a-1)\sinh^{-2}\biggl(\frac{\alpha x_-}{2}
\biggr) + 4a^2\alpha^2 +\nonumber\\
 &+&A \biggl[\exp(-2\alpha
x_1)-2 \exp(-\alpha x_1) + \exp(-2\alpha x_2)-2 \exp(-\alpha x_2)\biggr];
\nonumber\\
V(\vec x)&=& \alpha^2a(2a+1)\sinh^{-2}\biggl(\frac{\alpha x_-}{2}\biggr) +
4a^2\alpha^2 +\nonumber\\
 &+&A \biggl[\exp(-2\alpha
x_1)-2 \exp(-\alpha x_1) + \exp(-2\alpha x_2)-2 \exp(-\alpha x_2)\biggr],
\label{morse}
\ea
where $A$ is an arbitrary positive constant,
and $a$ is a parameter originating from the new
multiplicative constant
mentioned above. We will show in the next Sections that the range of variation
of this parameter $a$ will characterize the dynamics of the model.
It is perhaps interesting to remark that the reflection $a\to -a$
signals the supertransformation $Q^{+}\leftrightarrow Q^{-}$ and
$H\leftrightarrow \tilde H.$
Both potentials $V(\vec x)$ and $\tilde V(\vec x)$ are also invariant
under the interchange $x_1 \leftrightarrow x_2$ ("$x_{-}$-parity"
conservation). As for standard $P$-symmetry, this invariance leads to
classification of eigenfunctions according to their "$x_{-}$-parity"
values.

One easily recognizes in (\ref{morse}) a sum of two Morse potentials
plus a hyperbolic singular term
which prevents to apply the method of separation of variables for the system
(\ref{morse}). These singular terms in $\tilde V(\vec x), V(\vec x)$
can be both
attractive,
for the case $|a| > \frac{1}{2},$ or one repulsive and one attractive,
for the case $|a| < \frac{1}{2}.$ The parameter $a$
will be further constrained by the condition that the strength of the
attractive singularity at $x_-\to 0$ should not exceed the well known bound
$-1/(4x_-^2).$

\subsection*{\bf \quad 4.2.\quad Zero modes of $Q^+$.}
\hspace*{3ex}
The reason for the choice of
the model (\ref{16}) - (\ref{morse}) is that the corresponding
equations  for the zero modes
(\ref{etarho}) of $q^+$ can be solved analytically  in this case.
In particular
\cite{landau},
for the discrete spectrum
$\epsilon_n < 0$ the normalizable eigenfunctions are:
\be
\omega_n(\vec x) = \exp(-\frac{\xi_1+\xi_2}{2}) (\xi_1\xi_2)^{s_n}
F(-n, 2s_n +1; \xi_1) F(-n, 2s_n +1; \xi_2),
\label{hyper}
\ee
where $F(-n, 2s_n +1; \xi) $ is the standard degenerate (confluent)
hypergeometric function, reducing to a polynomial for integer $n,$ and
\ba
\xi_i&\equiv &\frac{2\sqrt{A}}{\alpha} \exp(-\alpha x_i); \label{xi}\\
s_n&=&\frac{\sqrt{A}}{\alpha}-n-\frac{1}{2} > 0; \label{si}\\
\epsilon_n&=&-A\biggl[1-\frac{\alpha}{\sqrt{A}}(n+\frac{1}{2})\biggr]^2.
\label{epsilon}
\ea
The number $(N+1)$ of normalizable zero modes (\ref{hyper}) is
determined by the inequality (\ref{si}).
As a final remark, let us point out that the zero modes $\omega_n(\vec x)$
of $q^+$ are entirely based on the "Morse-part" of the potentials, and
consequently do not depend on the parameter $a,$ which reflects the
strength of the singular part.

\subsection*{\bf \quad 4.3.\quad Normalizability of zero modes of $Q^+$.}
\hspace*{3ex}
We want to discuss the normalizability of the zero modes $\Omega_n(\vec x)$
which are connected with (\ref{hyper}):
\be
\Omega_n(\vec x) = \exp(\chi (\vec x)) \omega_n(\vec x).
\label{norm}
\ee
While in principle one can investigate the most general conditions for which
$\Omega_n(\vec x)$ is normalizable but  $\omega_n(\vec x)$ is not,
we will restrict ourselves for simplicity to the conditions for which
the normalizability of $\omega_n(\vec x)$ implies that for
$\Omega_n(\vec x).$ A constructive discussion of these restrictions will
be given in the following.

>From (\ref{chi}) and (\ref{cpm}) one can find the analytical expression:
\be
\exp(\chi (\vec x)) =
\exp(-a\alpha x_+)\vert\sinh \biggl(\frac{\alpha x_-}{2}\biggr)\vert^{-2a}=
\biggl(\frac{\alpha}{\sqrt{A}}\cdot\frac{\xi_1\xi_2}{|\xi_2 -\xi_1|}\biggr)
^{2a};
\quad \alpha >0.
\label{exp}
\ee
For $\xi_1\to\xi_2$ the term  $|\xi_2 -\xi_1|
^{-2a}$ in (\ref{exp}) requires for the normalizability of (\ref{norm})
that $a<1/4.$

At infinity (in the $\xi_1 \geq 0,\xi_2\geq 0$ quadrant)
the (\ref{exp}) does not
change essentially the behaviour of
$\omega_n (\vec x)$, which is $\exp(-\frac{\xi_1+\xi_2}{2}).$

At the origin (again in $\xi_1,\xi_2$) the behaviour of
$\omega_n (\vec x) \sim (\xi_1\xi_2)^{s_n},\quad s_n>0$ combines
with behaviour
at the origin of (\ref{exp}). In polar coordinates $$\xi_1=\xi\cdot\cos\phi ;
\quad \xi_2=\xi\cdot\sin\phi$$ the relevant integrand of
$\Vert\Omega (\vec x)\Vert^2$
at the origin reads:
$$
d^2x\,\Omega_n^2(\vec x)\sim \frac{d\xi}{\xi}\cdot\frac{d\phi}{\sin(2\phi )}
\cdot
\xi^{4(a+s_n)}
\cdot\frac{\biggl(\sin(2\phi)\biggr)^{2(2a+s_n)}}
{\vert\sin (\phi - \frac{\pi}{4})\vert^{4a}}.
$$

Combining all these restrictions, normalizability of $\Omega_n(\vec x)$
therefore is ensured by:
$$
s_n + a >0;\quad s_n + 2a >0;\quad s_n > 0;\quad
a < \frac{1}{4} .
$$

In addition, one has to impose the constraint
that the singularity for the Superhamiltonian
(i.e. both for $H$
and $\tilde H$) should be repulsive or, if attractive, should be
bounded, as explained at the end of Subsection 4.1, by
$4a(2a\pm 1)\geq -1/4.$
All these constraints can be implemented contextually leading to three
parameter families of models:
\be
a \in (-\infty,\, -\frac{1}{4}-\frac{1}{4\sqrt{2}});
\quad
s_n = \frac{\sqrt{A}}{\alpha}-n-\frac{1}{2} > -2a >0.
\label{region1}
\ee
Inequalities (\ref{region1}) can be satisfied by appropriate
choice of parameters $a$ and $A,$ and/or by suitable restriction of
the considered number $N$ of zero modes $\Omega_n(\vec x).$
It is perhaps interesting
to remark that the reflection symmetry $H\leftrightarrow \tilde H,\,
Q^+\leftrightarrow Q^-$ for $a\to -a$ is
broken by the fact that the operator $Q^-$ has no normalizable zero modes
when the parameters range in (\ref{region1}),
while $Q^+$ has. An alternative possible range of parameters which allows
to implement explicitly the reflection symmetry will be discussed in Section 6.

\subsection*{\bf \quad 4.4.\quad Algebraic solution.}
\hspace*{3ex}
In this Subsection we will prove explicitly that the operator $h,$
when acting in the linear space of the zero modes of $q^+,$ leaves
this space invariant.
Inserting (\ref{hyper}) into expression (\ref{eigen}) one obtains:
\be
h\omega_n(\vec x)=
-2(2a\alpha^2 s_n - \epsilon_n)\omega_n(\vec x) +\frac{4a\alpha^2 n}{2s_n+1}
\cdot\frac{(\xi_1\xi_2)^{s_n+1}}{(\xi_2-\xi_1)} \cdot\exp(-\frac{1}{2}
(\xi_1+\xi_2))
\cdot\Phi(-n+1,\,2s_n+2;\xi_1,\xi_2)\label{nonumber}
\ee
where the function $\Phi(b,c;\,\xi_1,\xi_2)$ is defined by:
$$
\Phi(b,c;\,\xi_1,\xi_2)\equiv [F(b,c;\xi_1)F(b-1,c-1; \xi_2)-
F(b,c;\xi_2)F(b-1,c-1; \xi_1)].
$$
Relations between contiguous hypergeometric functions lead to:
\ba
\Phi(b,c;\,\xi_1,\xi_2) =&& \frac{b-c}{(c-1)c}(\xi_2 -\xi_1)
[F(b,c;\xi_1)F(b,c+1; \xi_2)+\nonumber\\
 &&+\frac{b-c}{(c-1)c}\cdot\frac{b}{c(c+1)}\cdot
\xi_1\xi_2 \Phi(b+1,c+2;\,\xi_1,\xi_2).
\label{contig}
\ea

By making repeated use of (\ref{contig}) for $b=1-n$ one arrives after
laborious manipulations to:
\be
\Phi(1-n,c;\,\xi_1,\xi_2)=(\xi_2-\xi_1)\Sigma_{k=0}^{n-1} a_{nk}
(\xi_1\xi_2)^k\cdot [F(1-n+k,c+2k+1;\xi_1)F(1-n+k,c+2k+1; \xi_2)],
\label{sigma}
\ee
where
\ba
&&a_{n0}=-\frac{n+c-1}{(c-1)c};\quad a_{nk}=0 \,\,\,
for \quad k>n ; \nonumber\\
&&a_{nk}=-\frac{(n-1)!}{k!}\cdot
\frac{(n+c-1)(n+c)...(n+c+n-k-2)}{[(c-1)(c+2n-2k-2)]\cdot
[c(c+1)...(c+2n-2k-3)]^2} ,\,\, for \,\, k<n.\nonumber
\ea
In our case $c=2s_n+2$ and (\ref{sigma}) becomes:
$$
\Phi(1-n,2s_n+2;\,\xi_1,\xi_2)=(\xi_2-\xi_1)\Sigma_{k=0}^{n-1} a_{nk}
(\xi_1\xi_2)^{n-k-1}\, F(-k,2s_k+1;\xi_1)F(-k,2s_k+1; \xi_2).
$$
Inserting (\ref{sigma}) into (\ref{nonumber}) and taking into account that
$s_k-s_n=n-k$ (see (\ref{si})), an equation of the type
(\ref{matrixC}) is obtained:
\be
h\omega_n(\vec x) = \Sigma_{k=0}^N c_{nk}\omega_k(\vec x) =
-2(2a\alpha^2 s_n -\epsilon_n) \omega_n(\vec x) +
\frac{4a\alpha^2 n}{2s_n+1}\Sigma_{k=0}^{n-1} a_{nk}\omega_k(\vec x),
\label{triagonal}
\ee
showing that the matrix
$\hat C$ is triangular. We recall (see (\ref{norm})) that the zero
modes $\Omega_n (\vec x)$ are
related to $\omega_n (\vec x)$ by the similarity transformation.

\subsection*{\bf \quad 4.5.\quad Eigenfunctions of $H.$}
\hspace*{3ex}
The triangular matrix $\hat C$ with all different and nonzero
diagonal elements
can be diagonalized by the similarity transformation
(\ref{diag}) and its eigenvalues coincide
with the diagonal elements
$c_{kk} .$ From (\ref{diagonal}) one can construct a set of eigenfunctions
$\psi_n(\vec x)$ and
$\Psi_n(\vec x)$ of $ h$ and $ H,$
provided  matrix $\hat B$ is obtained
(see Eqs.(\ref{diag}), (\ref{diagonal})):
\be
\psi_{N-n}(\vec x) = \Sigma_{l=0}^N b_{nl}\omega_l(\vec x);\qquad
\Psi_{N-n}(\vec x) = \Sigma_{l=0}^N b_{nl}\Omega_l(\vec x).
\label{psiomega}
\ee
The index giving the numeration of the wave functions $\psi$ will be
elucidated below. The eigenvalues of $H$ which correspond to
$\Psi_{k}(\vec x)$ are expressed in terms of the parameters of the problem:
\be
E_{k}=c_{kk}=\lambda_{N-k}=-2(2a\alpha^2 s_{k}-\epsilon_{k}).
\label{energy}
\ee
>From (\ref{energy}) we conclude that assumptions made for the diagonal elements
of $\hat C$ are not very restrictive
due to the interplay of the parameters of the problem.

We start from the formal solution for the matrix elements
of $\hat B:$
\be
b_{m,p}=b_{m,N-m}\cdot\biggl[\Sigma_{l=1}^{N-p-1}
\biggl(\tau^{(m)}\biggr)^l\biggr]_{N-m,p},
\label{final}
\ee
where the $(N+1)$ triangular matrices labelled $\tau^{(m)},\,\, m=0,1,...,N$
are defined via the matrix elements of $\hat C:$
$$
\tau^{(m)}_{n,k}\equiv \frac{c_{n,k}}{c_{N-m,N-m}-c_{k,k}}.
$$
We stress that in (\ref{final}) the expression $\biggl(\tau^{(m)}\biggr)^l$
means the $l$-th power of the matrix $\tau^{(m)}.$ The repeated
index $(N-m)$ is frozen in (\ref{final})
and not summed over. This expression
allows to write all elements of the $m$-th row $b_{m,p}$ in terms
of the matrix $\tau^{(m)}$ and the arbitrary  value of the
element $b_{m,N-m}$
on the {\it crossed} diagonal. These arbitrary values can be fixed by the
normalization condition for the wave functions
$\Psi_{N-n}(\vec x)$ in (\ref{psiomega}).

>From the triangularity of $\hat C$ and $\tau^{(m)}$
in (\ref{final}) it follows that non-zero elements $b_{mp}$ are obtained
only for
$m+p\leq N$. This means the matrix $\hat B$ vanishes below the {\it crossed}
diagonal. It is now clear that the index of the wave function in
(\ref{psiomega})
is taken in a way to make $\Psi_k(\vec x)$ a linear combination of
the first $(k+1)$ zero modes $\Omega_l(\vec x);\,\, l=0,1,...,k.$ In particular,
$\Psi_0(\vec x) \sim \Omega_0(\vec x).$

We sketch a constructive algorithm to justify the result
(\ref{final}) for the matrix $\hat B$. It is not difficult to see that
indeed the system of equations
$$
\Sigma_{k=0}^N b_{i,k}c_{k,l}=\lambda_ib_{i,l}
$$
is solved for $i=0,\,\,l=N$ by (\ref{final}) with $\lambda_0=c_{N,N}$ and
arbitrary $b_{0,N}\neq 0$ due to the triangularity of $\hat C.$
One can also check that the solution holds for $i=0$ and $l=(N-1),(N-2),...,0$
by solving iteratively the corresponding linear equations.
For $i=1$ one convinces oneself that the first equation with our assumptions
implies $b_{1,N}=0,$ which signals the crossed triangularity mentioned above.
Following the steps as before $l=(N-1),(N-2),...,0$ one again can solve
iteratively with arbitrary element $b_{1,(N-1)}.$ Similar manipulations
can be performed for higher values $i=2,3...,N.$

All these wave functions $\Psi_k(\vec x)$ live in the space of zero
modes, in the next Subsection
we will construct also additional eigenfunctions, not linear
combinations of zero modes.

\subsection*{\bf \quad 4.6.\quad Additional eigenfunctions of $H.$}
\hspace*{3ex}
The eigenfunctions of the previous Subsection may be used for
constructing more general eigenfunctions of $h$ and of $H$
via a product ansatz:
\be
\phi (\vec x)\equiv \psi_0(\vec x)\cdot\Theta (\vec x); \qquad
\Phi (\vec x)\equiv \Psi_0(\vec x)\cdot\Theta (\vec x).
\label{product}
\ee
The eigenvalue equation for $h$ leads to:
\be
L\Theta (\vec x)=\gamma\Theta (\vec x),
\label{theta}
\ee
where
$$
L=-\alpha^2\xi_i^2\partial_{\xi_i}^2 +
\alpha^2\xi_i(\xi_i - 2s_0 -1)\partial_{\xi_i}-
4a\alpha^2\frac{\xi_1\xi_2}{\xi_2-\xi_1}(\partial_{\xi_1}-\partial_{\xi_2}).
$$
The sum over index $i$ is implicit. More specifically,
the eigenvalue equation
$$
h\phi (\vec x)=c_{00}\phi (\vec x)+
\psi_0(\vec x)L\Theta (\vec x)
$$
determines the new eigenvalues
\be
H\Phi (\vec x)=(E_0 +\gamma)\Phi (\vec x).
\label{values}
\ee

For the spectral problem (\ref{theta}) we can provide only particular
solutions by choosing suitable ansatzes for the functions $\Theta (\vec x)$.
It is useful to change variables:
$$
z_1 = \frac{1}{\xi_1}+\frac{1}{\xi_2};\quad
z_2 = \frac{1}{\xi_1}-\frac{1}{\xi_2},
$$
so that
$$
L=-\frac{\alpha^2}{2}\biggl[ (z_1^2+z_2^2)(\partial_{z_1}^2+\partial_{z_2}^2)
+ 4z_1z_2\partial_{z_1}\partial_{z_2} +
4\partial_{z_1}-2(2s_0-1)(z_1\partial_{z_1}+z_2\partial_{z_2})-
4a(2z_1\partial_{z_1}+z_2\partial_{z_2}
+\frac{z_1^2}{z_2}\partial_{z_2})  \biggr].
$$
The action of $L$ on monomial products $z_1^{\beta_1}\cdot z_2^{\beta_2}$
is:
\be
L (z_1^{\beta_1}\cdot z_2^{\beta_2})=-\frac{\alpha^2}{2}
\biggl[ \sigma (\beta_1,\beta_2) +\frac{z_1^2}{z_2^2} (\beta_2 -1
-4a)\beta_2 + \frac{z_2^2}{z_1^2}\beta_1(\beta_1-1)
+ \frac{4\beta_1}{z_1} \biggr](z_1^{\beta_1}\cdot z_2^{\beta_2}),
\label{monomial}
\ee
where constants $\sigma(\beta_1,\beta_2)$ read:
$$
\sigma (\beta_1,\beta_2)\equiv \beta_1(\beta_1-1)+\beta_2(\beta_2-1)
+4\beta_1\beta_2-4a(2\beta_1+\beta_2)-2(2s_0-1)(\beta_1+\beta_2).
$$

Using (\ref{monomial}) and linear combinations of two terms (\ref{monomial})
for
different powers $(\beta_1,\beta_2)$ and $(\tilde\beta_1,\tilde\beta_2),$
one can construct  only three solutions of (\ref{theta}):
\ba
&&1)\quad \Theta^{(1)}(\vec x) =
z_2^{(4a+1)};\quad\quad\quad\quad\quad\quad\quad\quad\quad\quad
\gamma^{(1)}= \alpha^2 (2s_0-1)(4a +1);
\label{a}\\
&&2)\quad
\Theta^{(2)}(\vec x) = z_2^{(4a+1)}
\biggl( z_1+\frac{2}{4a-2s_0+3} \biggr);\quad
\gamma^{(2)}= 4\alpha^2 (s_0-1)(2a + 1);
\label{bb}\\
&&3)\quad \Theta^{(3)}(\vec x) =
z_1-\frac{2}{4a+2s_0-1};\quad\quad\quad\quad\quad
\gamma^{(3)}= \alpha^2 \biggl(4a + 2s_0-1\biggr).
\label{c}
\ea

The eigenvalues $\gamma$ of (\ref{theta}) can be easily identified, when
the corresponding functions $\Theta (\vec x)$ do not spoil the normalizability
of $\Psi_0(\vec x).$
Within the bounds imposed (\ref{region1}) {\bf only the case (\ref{c}) is
acceptable} provided one also imposes $s_0>-a+\frac{1}{2}.$

With a similar procedure one can try to construct further eigenfunctions
of $H$ based on other $\Psi_k(\vec x);\,k>0$ in turn formed via
superpositions (\ref{psiomega}) of zero modes. Of course, this task will be more difficult than
the one we illustrated above, since the coefficient functions of the operator
$L$ will contain explicitly the hypergeometric functions from (\ref{hyper}).
In the particular case $n=0$, described above, the hypergeometric functions
reduce to $1.$

\section*{\bf 5.\quad Shape invariance for two-dimensional systems.}
\hspace*{3ex}
In Section 4 we studied the two dimensional spectral problem starting
from the linear space
spanned by zero modes $\Omega_n(\vec x)$ of supercharge $Q^+$ and then
constructing an additional eigenstate (\ref{c})
outside this space by choosing  a suitable ansatz. It is obvious that
algebraic methods to extend this construction to larger space of
eigenfunctions are highly welcome. In this respect the well known
(in one dimension) method of shape invariance (Subsection 5.1) is very
suggestive.
While in one dimension shape invariance amounts effectively to
exact solvability, in two dimensions we will show that one can achieve
partial solvability only.
The reasons of this will be particularly clear from an analysis of
simple two-dimensional
systems with separation of variables (Subsection 5.2). On one side, one
knows how to solve
the problem for this system by standard shape invariance for each degree
of freedom. On the other side, we also will consider this problem directly
from  a two-dimensional point of view and show that, in general, only
partial solvability will be achieved in this last approach. Finally
(Subsection 5.3), we will extend our method to the two-dimensional systems
with $SUSY-$separation of variables, already
described in Section 4, where only
partial solvability holds.

\subsection*{\bf \quad  5.1. \quad One-dimensional shape invariance and
solvability.}
\hspace*{3ex}
For reader's convenience we write the basic steps of standard shape
invariance in {\it absence of spontaneous SUSY breaking} \cite{review},
\cite{shape}.
One refers to shape invariance when a one-dimensional Superhamiltonian
$\hat H$ depends on a
parameter $a$ and, in addition, its components $H$ and $\tilde H$
satisfy:
\be
\tilde H(a)=H(\bar a) + {\cal R}(a),
\label{shape1}
\ee
where $\bar a=\bar a(a)$ is some new value of parameter,
which depends on $a,$ and $ {\cal R}(a)$ is a ($c$-number) function of $a.$
The absence of spontaneous breaking of supersymmetry for all values of $a$
implies that the lowest eigenvalue $E_0(a)$ of $H(a)$ vanishes and the corresponding
eigenfunctions $\Psi_0(a)$ are normalizable zero modes of $Q^+(a).$
It is well known \cite{review}, \cite{shape}, that the intertwining relations
\be
Q^-\tilde H(a)=H(a)Q^-(a)
\label{intertww}
\ee
with the standard first order supercharge (\ref{superch1})
allow in this case to solve the entire spectral problem for $H(a).$

The crucial steps are as follows. Start from
\be
H(\bar a)\Psi_0(\bar a)= E_0(\bar a)\Psi_0(\bar a)=0.
\label{sch}
\ee
Consider the relation (\ref{shape1}) to obtain:
\be
\tilde H(a)\Psi_0(\bar a)={\cal R}(a)\Psi_0(\bar a).
\label{tildesch}
\ee
It is important to remark that $\Psi_0(\bar a)\equiv \tilde\Psi_0(a)$
has no nodes and therefore is the ground state wave function of $\tilde H(a).$
The combination of (\ref{intertww}) and (\ref{tildesch}) yields:
\be
H(a)\biggl[ Q^-(a)\Psi_0(\bar a)\biggr]=
{\cal R}(a)\biggl[ Q^-(a)\Psi_0(\bar a)\biggr].
\label{R}
\ee
Provided
$\biggl[ Q^-(a)\Psi_0(\bar a)\biggr]$ is normalizable,
we have generated an excited state of $H(a),$ and thus
${\cal R}(a)$ is naturally positive. It is clear that these steps can be
repeated up to
the last step, where the resulting wave function $\Psi$ will no more be
normalizable. There are notorious cases (oscillator-like potentials) where
the spectrum is not bounded from above.

It is also clear that the isospectrality of $H(a)$ and $\tilde H(a)$
(up to
the only zero mode $\Psi_0(a) )$ implies that there is no eigenvalue of $H(a)$
between zero and the ground state energy $\tilde E_0(a)$ of $\tilde H.$
This observation leads to a proof that after suitable iterations one gets
the entire spectrum of $H(a).$
This method is referred as algebraic solvability (or complete solvability)
by shape invariance in one-dimensional SUSY QM.

\subsection*{\bf \quad 5.2.\quad Two-dimensional shape invariance for
systems with separation of variables: solvability or partial solvability.}
\hspace*{3ex}
Already the trivial two-dimensional model with separation of variables:
$$
H(\vec x)=H_1(x_1)+H_2(x_2);\quad \vec x=(x_1,x_2)
$$
shows that there is
a considerable difference with respect to the one-dimensional case. The crucial reason is
that the space of zero modes of supercharges becomes now of higher
dimensionality including the products of one-dimensional zero modes
of the first Hamiltonian times
all states of the second Hamiltonian and vice versa.

In order to realize a nontrivial intertwining relations one can consider
factorized supercharges of second order written as products of first order
supercharges\footnote{Note that in such models $H$ allows the separation of
variables, but $Q^{\pm}$ do not. If one wants also $Q^{\pm}$ to allow the
separation of variables, then one would have to consider $Q^{\pm}=Q^{\pm}_i.$}:
\be
Q^{\pm}=Q_1^{\pm}\cdot Q^{\pm}_2; \quad Q^{\pm}_i= (\mp\partial_i + W_i(x_i)).
\label{sepcharge}
\ee
Now suppose that $H_1$ and $H_2$ both are shape invariant:
$$
\tilde H_1(a_1)=H_1(\bar a_1)+{\cal R}_1(a_1);\quad
\tilde H_2(a_2)=H_2(\bar a_2)+{\cal R}_2(a_2),
$$
i.e. $H$ is shape invariant with a vector parameter ${\bf a}=(a_1,a_2):$
$$
\tilde H({\bf a})=H({\bf \bar a})+{\cal R}({\bf a}).
$$

While iterations analogous to the previous Subsection are obviously possible,
it is clear that one can not argue about the entire solvability of
the spectral problem, because in general many zero modes of (\ref{sepcharge})
exist. Their number depends on the confining properties of $H_1$ and $H_2.$
For example, in a case of oscillator-like potentials this number becomes
infinite, and they are distributed over the whole spectrum.
In this case only partial solvability of $H$ can be achieved by the choice
of (\ref{sepcharge}) and shape invariance.
Of course, one can solve such trivial models by separate use of
$Q^{\pm}=Q^{\pm}_i,$ which allows to solve the entire spectrum of
two-dimensional model in terms of the one-dimensional ones.

\subsection*{\bf \quad 5.3. Shape invariance and partial
solvability for two-dimensional systems.}
\hspace*{3ex}
Let us suppose to have two-dimensional system with a Hamiltonian $H,$ which
is related to $\tilde H$ by (\ref{shape1}). For simplicity,
we have assumed\footnote{We recall that, in general, there is no
connection between
the dimensionality of the Schr\"odinger equation and the dimensionality of
the parameter manifold.}
that shape invariance is realized only with the parameter $a.$
Two-dimensional SUSY
QM, realized via (\ref{intertw2}) - (\ref{superch22}), does not
identify zero modes
of $Q^{\pm}$ with the ground state of the Hamiltonian. Thus one has to
repeat the steps (\ref{shape1}) - (\ref{R}) of Subsection 5.1
by taking into account $E_0(a)\neq 0.$
In order to make our discussion more explicit we will from now on
refer explicitly to the model (\ref{cpm}) - (\ref{morse}) with the
parameter $a$ being bound to (\ref{region1}), as described in the Section 4.

First of all we observe that this model is indeed shape invariant with
\be
\bar a = a-\frac{1}{2};\quad {\cal R}(a)=\alpha^2(4a-1).
\label{bara}
\ee
We remark also that the infinite domain given by (\ref{region1}) allows iterations
of (\ref{shape1}). The starting point is to write (\ref{R}):
\be
H(a)\biggl[ Q^-(a)\Psi_0(a-\frac{1}{2}) \biggr] = \biggl(E_0(a-\frac{1}{2})
+ {\cal R}(a)\biggr)\cdot\biggl[ Q^-(a)\Psi_0(a-\frac{1}{2}) \biggr],
\label{RR}
\ee
where $E_0(a)$ and $\Psi_0(a),$ not to be identified with ground state,
are given by (\ref{psiomega}) and (\ref{energy}). Thus we have constructed the
new eigenstate and eigenvalue of $H(a),$ provided $Q^-(a)\Psi_0(a-\frac{1}{2})$
is normalizable. We notice that the eigenvalue $\biggl(E_0(a-\frac{1}{2}) +
{\cal R}(a)\biggr)$ is larger than $E_0(a)$ with the bounds of (\ref{region1}).

It is interesting to compare this first iteration (\ref{RR}) of shape
invariance  with the solution (\ref{c}) obtained
in the framework of the ansatz described in Subsection 4.6.
Their eigenvalues coincide {\bf precisely} :
$$
E=E_{0}(a)+\gamma^{(3)}(a)=E_{0}(a-\frac{1}{2})+ {\cal R}(a)=
\alpha^{2}[4a(1-s_{0})+(2s_{0}-1)]+2\epsilon_{0},
$$
while the eigenfunctions (both vanishing for $ x_{-}=0 $) differ by a
factor $ x_{-} / |x_{-}| $ reflecting the opposite values of
"$ x_{-}$-parity".

The next iteration of shape invariance  will give:
\be
H(a)\biggl[ Q^-(a)Q^-(a-\frac{1}{2})\Psi_0(a-1) \biggr] =
\biggl(E_0(a-1)
+ {\cal R}(a-\frac{1}{2}) + {\cal R}(a)\biggr)\cdot
\biggl[ Q^-(a)Q^-(a-\frac{1}{2})\Psi_0(a-1) \biggr],
\label{RRR}
\ee
and the new eigenfunction $ Q^-(a)Q^-(a-\frac{1}{2})\Psi_0(a-1) $ can be
written explicitly as function of $ \vec x . $
Provided normalizability is ensured, one can thereby construct a chain
by successive iterations of (\ref{RR}) and (\ref{RRR}), since $Q^-(a)$
has no normalizable zero modes in (\ref{region1}).
The end point of such a chain will be given by non-normalizability
of the relevant wave function.

\section*{\bf  6.\quad Discussions and conclusions.}
\vspace*{0.2cm}
\hspace*{3ex}
We want to point out that $SUSY-$separation of variables can be implemented
completely independently from shape invariance. To this end we present
a model where $SUSY-$separation of variables holds but shape invariance does not
apply.

Consider the model (\ref{cpm}) - (\ref{morse}) with a choice for
parameter $a$ { \bf alternative} to (\ref{region1}):
\be
a \in ( -\frac{1}{4} ,\,\frac{1}{4} );
\quad s_0>2(|a|+1).
\label{region2}
\ee
It is obvious that shape invariance does not apply since the domain
(\ref{region2}) is too small.
One important property of this model is that the reflection symmetry $a \to -a$
is explicitly implemented.  This allows normalizability
of both zero modes\footnote{We recall that $\omega_n(\vec x)$
by construction do not depend on the parameter $a$ (see Subsection 4.2).}
$\tilde\Omega (a)=\Omega (-a)$ of $Q^-$ and $\Omega (a)$
of $Q^+$ and also of {\bf all three} wave functions
$\Phi^{(i)}(\vec x)$ with $\Theta^{(i)}(\vec x)$ given in (\ref{a}) --
(\ref{c}).

Although we have achieved only partial solvability, we now
present a short discussion of the quantum integrals of motion
({\bf symmetry operators}), which exist \cite{HSUSY}, \cite{david},
\cite{classical} for all HSUSY QM systems.
Indeed, the intertwining relations (\ref{intertw2}) lead to existence
of the symmetry operators $ \tilde R, R $ for the Hamiltonians
$\tilde H,\,H,$ correspondingly:
\be
\tilde R = Q^+Q^-,\quad R = Q^-Q^+,\quad
[R, H] = 0, \quad [\tilde R, \tilde H]=0.\label{symm}
\ee
In 1-dim case \cite{HSUSY} these quantum integrals of motion
$ R, \tilde R $ become polynomials of $ H, \tilde H $ with constant
coefficients. The distinguishing peculiarity of two-dimensional case is
given by \cite{david}, \cite{classical}  nontrivial symmetry operators
$ \tilde R, R $ which are not reduced to functions of the Hamiltonians
$ H, \tilde H,$
i.e. all two-dimensional systems (\ref{potential}) which solve the
intertwining relations (\ref{intertw2}) are {\bf integrable}.

More specifically, it was shown in \cite{david}, in the case of second order
HSUSY, that for the particular case of the unit metrics
$ g_{ik}=\delta_{ik} $ the fourth order
operators $ \tilde R, R $ can be written
as second order differential operators up to a function of
$ H, \tilde H $ (R-separation of variables \cite{miller}).
For all other metrics $ g_{ik} $
the operators $ \tilde R, R $ are of fourth order in derivatives.

By construction, the quantum integral of motion
$ R=Q^{-}Q^{+}$ gives zero when acting onto the eigenfunctions
of Hamiltonian $ \Psi_{k} $ from (\ref{psiomega}). A direct calculation
shows that the additional eigenfunctions $ \Phi^{(i)}(\vec x) $ from (\ref{a}),
(\ref{bb}), (\ref{c}) are also eigenfunctions of operator $ R :$
$$
R\Phi^{(i)}(\vec x) =\alpha^{2} r_{i}\Phi^{(i)}(\vec x)
$$
with corresponding eigenvalues:
\ba
r_{1}=-(4a+1)(2s_{0}-1)(4a-2s_{0}+1),\nonumber\\
r_{2}=-16(2a+1)(s_{0}-1)(2a-s_{0}+1),\nonumber\\
r_{3}=-(4a-1)(2s_{0}-1)(4a+2s_{0}-1).\nonumber
\ea
Thus the eigenfunctions (\ref{psiomega})
and (\ref{a})-(\ref{c})
are part of the  system of common eigenfunctions of two
hermitian mutually commuting operators, the Hamiltonian $ H $ and the quantum
integral of motion $ R.$

In conclusion, we have formulated two new approaches to partial solvability
of two-dimensional quantum systems:

1) The $SUSY-$separation of variables method
(Section 3) is to be considered as a particular branch of the method of
separation of variables, though not for the Hamiltonian,
but for the supercharge.
It is in contrast to the more trivial case, where $H$ allows separation,
but $Q^{\pm}$
is factorized and does not allow for the separation (Subsection 5.2).

2) The method of shape invariance, well known for the one-dimensional SUSY QM,
has been reformulated (Subsection 5.3) for the two-dimensional systems, which
depend parametrically on $a,$ in the same way as for one dimensional
shape invariant systems. This construction is based on the knowledge of the
eigenfunctions $\Psi_0(a)$ and of the eigenvalues $E_0(a)$
(for $a$ in the domain (\ref{region1})). We have
illustrated this method applying (Subsection 5.3)
it to a "singular" two-dimensional Morse system (Subsection 4.1)
with Higher order SUSY QM in presence of a variety of zero modes of $Q^+.$

\section*{\bf Acknowledgements}
M.V.I. is indebted to INFN, the University of
Bologna, the Scientific Council of Norway and University of Helsinki
for the support and hospitality.
He also is indebted to Prof.I.Komarov
(S.-Petersburg) for illuminating advices and warmly acknowledges kind
hospitality and useful discussions with Prof.P.Osland (Bergen) and
Prof.M.Chaichian (Helsinki).
D.N.N. thanks GSB Company (President Prof.G.Sokhadze) for the
providing Internet access.
This work was partially supported by the Russian Foundation for Fundamental
Research (Grant No.02-01-00499).

\end{document}